\documentclass[10pt,leqno]{article} 
\usepackage{graphicx}
\usepackage{indentfirst,csquotes}

\usepackage{amssymb,amsthm,amsmath}
\usepackage{xcolor,paralist,hyperref,titlesec,fancyhdr,etoolbox}

\titleformat{\section}[display]
  {\normalfont\huge\bfseries\centering}
  {}
  {0pt}
  {\Large}
\titlespacing*{\section}{0pt}{0ex}{0ex}

\hypersetup{ colorlinks=true, linkcolor=blue, filecolor=black, urlcolor=black }

\usepackage{lipsum}
\usepackage{booktabs}
\usepackage{float}

\usepackage[printonlyused,smaller,withpage]{acronym}
\usepackage{dirtytalk}
\usepackage{multirow}

\usepackage{listings}

\definecolor{codegray}{gray}{0.95}
\definecolor{codecomment}{gray}{0.35}
\usepackage{tikz}
\usepackage{arydshln}

\begin{document}
\title{DCI: Dependency Confidence Index for Assessing Open-Source
Dependency Trustworthiness} 
\author{Clemens Albrecht$^1$ \and Stefan Reitmann$^{1,*}$}
\date{%
    $^1$Chemnitz University of Technology\\
    $^*$\href{mailto:stefan.reitmann@informatik.tu-chemnitz.de}{stefan.reitmann@informatik.tu-chemnitz.de}\\
    \vspace*{2em}
}
\maketitle


\begin{abstract}
Selecting trustworthy open source software dependencies remains a major challenge in software supply chain security. 
We present the \ac{DCI}, a composite formative index that combines nine empirically weighted trust factors into a single normalized composite score for dependency selection. 
\ac{DCI}’s trust factors combine insights from a systematic literature review and an exploratory \ac{AHP} survey of ten software developers, highlighting security, source code quality, and project health as the most influential dimensions. Following Goal-Question-Metric methodology, we implemented 12 automated measurements using SonarQube, GitHub APIs, and OpenSSF Scorecard data, deployed in a containerized evaluation platform.
We conducted a pilot evaluation of the normalized \ac{DCI} on 92 popular PyPI packages, observing moderate agreement with OpenSSF Scorecard scores and perfect test–retest reliability.
Analysis reveals process-based factors (dependency management, CI) dominate scores on high-quality packages, while security metrics saturate - suggesting \ac{DCI}'s complementary role to existing tools.
Our publicly available implementation provides a foundation for open source software trustworthiness research and practical dependency auditing.
\end{abstract} 

\section{Introduction}

Digital devices power most of our daily interactions - from self-service checkouts to mobile tickets.
The scale of modern software development necessitates third-party \ac{OSS} components, 
which together form complex supply chains where a single compromised dependency can undermine entire applications.
Recent audits show 96\% of 1,000 commercial projects contain \ac{OSS}, comprising 74\% of their codebases~\cite{ossra2024}.
Yet 84\% of these codebases harbor vulnerabilities, and 96\% of downloaded vulnerable components had fixes available~\cite{mayhew10thAnnualState}.
Malicious \ac{OSS} packages surged 156\% year-over-year to 704,102 in 2024~\cite{mayhew10thAnnualState}.

While metrics exist for individual quality signals (e.g., vulnerability counts, test coverage), developers lack a unified, interpretable score for dependency selection.
Popular packages address vulnerabilities 32\% faster~\cite{mayhew10thAnnualState}, yet developers remain uncertain which factors - security, code quality, process maturity, project health - matter most, or how to weight them.

This paper presents the \textbf{\acf{DCI}}, a composite formative index that combines 9 preliminary weights derived from an exploratory \ac{AHP} survey into a single, normalized score in the $[0,1]$ interval for \ac{OSS} dependency selection. Our contributions are:

\begin{itemize}
\item A trust model with 9 factors (security, source code quality, documentation completeness, license declaration, development process quality, project health, release cadence, dependency management, reputation) informed from systematic literature review and weighted via \acf{AHP} survey ($n=10$ developers).
\item 12 automated, normalized measurements implemented via SonarQube, GitHub \acp{API}, and OpenSSF Scorecard data, deployed in a containerized evaluation platform.
\item A pilot evaluation on 92 popular PyPI packages, in which normalized \ac{DCI} scores   show a moderate correlation with OpenSSF Scorecard ($\rho=0.396$, $p=0.0002$) and identical results under test–retest execution for a subset of packages.

\end{itemize}

Analysis reveals process signals (\ac{CI}, dependency management) dominate high-quality packages while security metrics saturate, positioning \ac{DCI} as complementary to existing tools.
Our publicly available implementation enables further \ac{OSS} trustworthiness research. Practically, \ac{DCI} is intended to support dependency review rather than to replace specialized security tools. A developer could use the index as an additional screening signal when comparing candidate libraries, while inspecting the individual factor values to identify weaknesses related to security, code quality, project maintenance, development process, dependency management, or licensing. The score is therefore intended to organize heterogeneous evidence and support further investigation, not to provide an automatic guarantee that a dependency is safe.

The remainder of this paper is structured as follows: Sec.~\ref{sec:background} discusses software supply-chain attacks, trust, trust transitivity, trust factors, and existing software trust solutions. Sec.~\ref{sec:concept} presents the conceptual construction of \ac{DCI}. Sec.~\ref{sec:methodology} describes the selected trust factors, measurements, weighting procedure, and system implementation. Sec.~\ref{sec:eval} reports the pilot evaluation on popular PyPI packages, including the comparison with OpenSSF Scorecard and the test-retest experiment. Sec.~\ref{sec:conclusion} discusses the findings, limitations, intended usage scenarios, and future research directions.

\section{Background}
\label{sec:background}

This section starts with an elaboration on software supply chain attacks. It then follows this with an examination of the definition of trust, and an exploration of the concept of trust transitivity. Lastly, it investigates central trust factors and software trust solutions.

\subsection{Software Supply Chain Attacks}

Software supply chain attacks target three vectors: dependencies, build infrastructure, and humans~\cite[pp.\ 2--3]{williamsResearchDirectionsSoftware2025}.
Public package repositories introduce numerous trust boundaries that attackers exploit~\cite{ohmBackstabbersKnifeCollection2020}.
While SolarWinds exemplified build server compromise via weak credentials~\cite[p.\ 44]{yangSolarWindsSoftwareSupply2022}, dependency poisoning and maintainer account takeovers remain prevalent.

This work focuses on dependencies and humans, echoing Ken Thompson's Turing Award observation: ``To what extent should one trust a statement that a program is free of Trojan horses. Perhaps it is more important to trust the people who wrote the software''~\cite[p.\ 761]{thompsonReflectionsTrustingTrust1984}.

\subsection{Trust}
\label{ss:trust}

Early social psychology literature defines trust as a ``willingness to take risks''~\cite[p.\ 1306]{johnson-georgeMeasurementSpecificInterpersonal1982}.
Johnson-George and Swap illustrate domain-specificity with a pet sitter trusted for cat care but not car repair, versus a mechanic trusted for vehicles but not pets~\cite{johnson-georgeMeasurementSpecificInterpersonal1982}.
Their survey using 9-point trust scales identified dependability and reliableness as key factors relevant to software~\cite[p.\ 3]{johnson-georgeMeasurementSpecificInterpersonal1982}.

Mayer et al.~\cite{mayerIntegrativeModelOrganizational1995} adapt this to organizational contexts, defining trust as ``the willingness of a party to be vulnerable to the actions of another party based on the expectation that the other will perform a particular action important to the trustor, irrespective of the ability to monitor or control that other party''~\cite[p.\ 4]{mayerIntegrativeModelOrganizational1995}.
This relational view distinguishes trusting software from trusting its producers.

To analyze \ac{OSS} trustworthiness precisely, we adopt Becerra et al.'s distinction: trust is the belief in the \textit{trustworthiness} of the trustee, where trustworthiness is a property of the trustee itself~\cite{becerraTrustworthinessRiskTransfer2008}.

\subsection{Trust Transitivity}

Research commonly assumes trust transitivity: if Alice trusts Trent and Trent trusts Bob, then Alice trusts Bob.
However, applying this to software trustworthiness requires validating it holds in context~\cite{josangPopeTransitivitySemantics}.

J{\o}sang and Pope~\cite{josangPopeTransitivitySemantics} formalize path-based transitivity: the final edge must be \textit{functional trust} (direct belief, e.g., Trent trusts Bob's car expertise), while prior edges can be \textit{referral trust} (recommendations) if authenticated.
Long paths reduce confidence; parallel paths increase it.
Liu et al.~\cite{liuTrustTransitivityComplex2011} extend Mayer et al.~\cite{mayerIntegrativeModelOrganizational1995} by requiring \textit{context} constraints - Alice trusts mechanic-Bob for cars, not programming.

We adopt this view: trust is transitive when context-respecting chains satisfy J{\o}sang and Pope's criteria, justifying \ac{OSS} dependency evaluation via maintainer/project signals.

\subsection{Trust Factors}
\label{ss:trustfactors}
Section \ref{ss:trust} defines trust to consist of a number of trust factors. In the context of software,
these factors are derived from functional and non-functional requirements. Literature
provides many methods for evaluating specific requirements. \cite{houSystematicLiteratureReview2023} and \cite{houSurveyStateoftheartApproaches2024} provide dimensions for classifying trust factors that are derived from requirements. \cite{houSystematicLiteratureReview2023} classifies trust factors as either intrinsic or extrinsic. Intrinsic factors “represent a product’s physical
attributes” \cite[p.~11]{houSystematicLiteratureReview2023}. Extrinsic factors “represent external attributes, such as the product’s reputation, cost, licenses, or capability of the software producer” \cite[p.~11]{houSystematicLiteratureReview2023}. \cite{houSurveyStateoftheartApproaches2024} defines a second dimension. Trust factors are classified based on the software being either close or open source as seen in \cite[Tab.~1]{houSurveyStateoftheartApproaches2024}. \cite{liExploringFactorsMetrics2022} investigates factors that developers evaluate when adopting a new open source software component. 

Notable in this list is architecture conformance. This means, developers are checking the
source code, and are determining if the source code matches the documentation provided.
Since \cite{liExploringFactorsMetrics2022} uses an empirical study, the factors in this list are factors, that can be easily obtained from looking at the repository of a project. \cite{houSystematicLiteratureReview2023} provides a meta analysis of the scientific literature. As a result, the factors closely match functional and non-functional requirements as used in software engineering.

Contrary to \cite{liExploringFactorsMetrics2022}, \cite{houSystematicLiteratureReview2023} includes security as a trust factor. Furthermore, it lists factors that have to be evaluated using task-specific tooling like static analysis tools. \cite{houSystematicLiteratureReview2023} also includes cost as a trust factor because the analysis in this study is not constrained to open source software. \cite{boughtonDecomposingMeasuringTrust2024} uses a different approach for establishing trust factors. They focus on the work of \cite{mayerIntegrativeModelOrganizational1995} and specify the factors. These factors are defined by \cite{boughtonDecomposingMeasuringTrust2024} as following: Propensity can be observed by measuring how quickly a maintainer of a project gives write access to the repository to a new contributor. Ability is the power of a new contributor to abuse the trust given to them. Benevolence can be measured by looking at the past history of the new contributor and judging, whether they have acted in good-faith in the past. Integrity evaluates, whether the new contributor has followed security best practices in the past.

This list of factors only evaluates the developers of the software. It does not include an analysis of the software itself. The differences between these methodologies show, that there is a considerable divergence between trust evaluation as proposed by scientific literature, and the actual methods employed by practitioners. Table \ref{tab:trust-factor-mapping} compares the different trust factor definitions presented. \cite{houSystematicLiteratureReview2023} is used as a reference definition because it is the result of a meta analysis. The table maps \cite{liExploringFactorsMetrics2022} and \cite{boughtonDecomposingMeasuringTrust2024} against this list. The comparison shows, that the definitions by \cite{liExploringFactorsMetrics2022} and \cite{boughtonDecomposingMeasuringTrust2024} can be fully represented by the definition in \cite{houSystematicLiteratureReview2023}. As a result, the definition of trust factors by \cite{houSystematicLiteratureReview2023} is used in this work.

\begin{table}[ht!]
\centering
\renewcommand{\arraystretch}{1.25}
\caption{Comparison of Trust Factor Definitions.}
\label{tab:trust-factor-mapping}
\begin{tabular}{|p{0.28\textwidth}|p{0.33\textwidth}|p{0.28\textwidth}|}
\hline
\textbf{Hou et al.~\cite{houSystematicLiteratureReview2023}} & \textbf{Li et al.~\cite{liExploringFactorsMetrics2022}} & \textbf{Boughton et al.~\cite{boughtonDecomposingMeasuringTrust2024}} \\
\hline
Security, Vulnerability, and Attack Proneness \textbf{SV} &
 &
Propensity, Integrity \\
\hline
Quality and Development Process \textbf{QP} &
Functionality, Reliability, Benchmarks/Tests &
 \\
\hline
Source Code Quality and Architecture \textbf{CQ} &
Architecture Conformance &
 \\
\hline
Versions and Dependencies \textbf{VD} &
 &
 \\
\hline
Structural Assurance \textbf{SA} &
3rd Party Assessment &
 \\
\hline
Documentation \textbf{DO} &
 &
 \\
\hline
Popularity \textbf{PO} &
Project Health &
 \\
\hline
Reputation \textbf{RE} &
Provider Reputation &
Ability, Benevolence, Integrity \\
\hline
Cost \textbf{CO} &
 &
Benevolence \\
\hline
\end{tabular}
\end{table}

\subsection{Software Trust Solutions}

This section compares existing software trust solutions. These are \ac{OpenSSF} Scorecard, SonarQube, and OSSGadget. For each solution, it explores how many of the trust factors defined in Section \ref{ss:trustfactors} can be evaluated using that solution. Lastly, this section compares these solutions and discusses the gaps presented between these solutions and the trust
factor definition.

\subsubsection{OpenSSF Scorecard}

The \ac{OpenSSF} Scorecard \cite{rinivasanOpenSSFScorecard2025} project is an initiative by the \ac{OpenSSF},
to provide \ac{OSS} projects with a tool, to help them improve their
security practices \cite[p.~1]{zahanOpenSSFScorecardPath2023}. To this end, Scorecard can be used to evaluate individual
GitHub repositories, but it also provides a database containing Scorecard scores for popular
\ac{OSS} projects \cite[p.~1]{zahanOpenSSFScorecardPath2023}.
Because this tool is developed by the \ac{OpenSSF}, it mostly focuses on security. It checks for
vulnerabilities, the use of a dependency update service, the use of fuzzing tools, and secure
management of the GitHub repository. Scorecard checks for process quality by checking
the use of \ac{CI} tests, and mandatory code reviews. However, it
does not directly evaluate code quality, as it only checks the use of static analysis tools.
Some tests on dependencies are performed, as it checks the use of dependency version
checking tools. It also evaluates popularity, reputation to an extent, by checking for the
participation of multiple contributors, and the existence of a special best practices badge.
Only the absence of costs are checked by looking for a license declaration. Both structural
assurances and documentation are not verified.
To evaluate the effectiveness of using Scorecard to improve security practices, \cite[p.~4]{zahanSoftwareSecurityPractices2023} used regression models to explore the relationship between high Scorecard scores, and
the vulnerability count. \cite{zahanSoftwareSecurityPractices2023} shows, that a high score corresponds to a higher number of vulnerabilities, as the study showed a “0.5-unit increase in vulnerability count for every
unit increase in aggregate security score” \cite[p.~8]{zahanSoftwareSecurityPractices2023}. This result is explained by the higher
popularity of projects with better security practices and the correspondingly increase in the scrutiny of these packages \cite[p.~9]{zahanSoftwareSecurityPractices2023}.

\subsubsection{SonarQube}

SonarQube is an \ac{ASAT} produced by SonarSource \cite{lancelotSonarQube2025}. It is one of the most widely used \acp{ASAT} in the context of \ac{CI} \cite{vassalloContinuousCodeQuality2018}. As a static analysis tool, SonarQube focuses mostly on security and code quality. It can fully evaluate these two trust factors. Security is evaluated by flagging the use of dangerous \acp{API} and known vulnerabilities like SQL injections. SonarQube evaluates code quality by measuring the size of the code, its complexity, and maintainability. Maintainability is measured using
technical debt and code smells.
SonarQube tests three trust factors only partially. Those are development process quality, and documentation. It checks the development process quality by measuring test coverage. Because it is an \ac{ASAT}, it does not evaluate the use of best practices regarding the use of source code management tools. SonarQube checks the existence of documentation by
counting amount of comment lines found in the code. It does not check actual project documentation. Due to the limited scope of the software, five trust factors cannot be analyzed using SonarQube.
While the use of \acp{ASAT} is not very common in the developer community \cite[p.~6]{vassalloContinuousCodeQuality2018}, when one is used, developers only fix some of the reported issues \cite[p.~8]{vassalloContinuousCodeQuality2018}. However, \cite{digkasCanCleanNew2022} shows that an improvement of code quality result from a use of static analysis tools reduces the accumulation of technical debt overt time.

\subsubsection{OSSGadget}

Similar to \ac{OpenSSF} Scorecard, OSSGadget is a collection of tools, built by Microsoft, which
can be used to evaluate the trustworthiness of an open source project \cite{stoccoOSSGadget2025}. Like Scorecard,
it focuses on security. It contains a static analysis tool, that flags notable code patterns in an application. Also, it contains a tool which attempts to detect malicious backdoors. However, it does not scan for vulnerabilities in the code. Notably, it includes an implementation of various project health metrics. It also contains a tool which tries to detect typosquatting
packages.

The two trust factors that are fully satisfied by this tool are popularity and reputation, both of which are implemented by the project health checker. Because it does not check for vulnerabilities, the security factor is only partly satisfied. The health checker also partly satisfies the process quality factor, however as OSSGadget does not check development practices like code testing, this factor is not fully satisfied. In contrast to Scorecard and SonarQube, OSSGadget has not been thoroughly evaluated. \cite{291044} examines only the typosquatting detector, finding that it does not detect a large proportion of possible package name pairs.

\subsubsection{Comparison}

While the tools that were evaluated in this section are not specifically made to evaluate
trust, the analysis has shown, that all three of them can be used to check a number of
trust factors. The results are shown in Table \ref{tab:trust-factors}. It shows that all three tools can evaluate
two factors fully (\textbullet), some partially (\(\circledcirc\)). Scorecard and SonarQube show very promising results for evaluating the security of a project, which OSSGadget can only partially fulfill. On the other hand,
OSSGadget implements many checks for evaluating the health of an \ac{OSS} project and therefore is able to evaluate the popularity and reputation factors.

\begin{table}[htbp]
\centering
\renewcommand{\arraystretch}{1.2}
\caption{Comparison of trust factors across \ac{OpenSSF} Scorecard, SonarQube and OSSGadget.}
\label{tab:trust-factors}
\begin{tabular}{|c|c|c|c|}
\hline
\textbf{Trust Factors} & \textbf{\ac{OpenSSF} Scorecard} & \textbf{SonarQube} & \textbf{OSSGadget} \\
\hline
\textbf{SV} & \textbullet & \textbullet & \(\circledcirc\) \\
\hline
\textbf{QP} & \textbullet & \(\circledcirc\) & \(\circledcirc\) \\
\hline
\textbf{CQ} & \(\circledcirc\) & \textbullet & - \\
\hline
\textbf{VD} & \(\circledcirc\) & - & - \\
\hline
\textbf{SA} & - & - & - \\
\hline
\textbf{DO} & - & \(\circledcirc\) & - \\
\hline
\textbf{PO} & \(\circledcirc\) & - & \textbullet \\
\hline
\textbf{RE} & \(\circledcirc\) & - & \textbullet \\
\hline
\textbf{CO} & \(\circledcirc\) & - & - \\
\hline
\end{tabular}
\end{table}

If the capabilities of all three programs are combined, the practitioner can evaluate all trust factors, except structural assurances. The lack of structural assurance evaluation can be explained by the rare use of software certification in the open source space. Since, almost no \ac{OSS} software is certified, it is fruitless to implement such a check. The problem
that arises from the use of all three tools is the significant work that has to go into the implementation of these tools into the development workflow. \cite{witscheyQuantifyingDevelopersAdoption2015} shows that the adoption of security tools by developers is hindered by the complexity of these tools. Therefore, the use of three tools to evaluate the trustworthiness of a software dependency would likely result in reduced adoption. Therefore, this work will implement a tool, that can evaluate important trust factors in one tool, and present developers with actionable insights to improve the adoption of such tooling into development workflows. 

The main advantage of \ac{DCI} is not that it contains more information than its component measurements. Rather, it provides a common decision-oriented representation of heterogeneous signals that are otherwise distributed across several tools. This may support initial dependency triage and make trade-offs more visible. The composite score must nevertheless be accompanied by its factor-level measurements, because a single score can conceal important weaknesses and cannot replace specialized security or quality analysis.

\section{Concept}
\label{sec:concept}

We construct the \ac{DCI} as a composite formative index that combines multiple trust factor measurements into a single, normalized trustworthiness score in a $[0,1]$ interval for \ac{OSS} libraries.
A value closer to 1 indicates a more favorable combination of the measured signals within the selected normalization scheme, whereas a value closer to 0 indicates less favorable measured signals. This scale is relative to the operational definitions and normalization bounds of the current implementation and must not be interpreted as a probability of safety or as an absolute trustworthiness level.
Unlike reflective scales (where indicators reflect a latent variable), formative indices measure how indicators causally influence the target variable~\cite[p.\ 183]{devellisScaleDevelopmentTheory2022}.

DeVellis~\cite[pp.\ 202--219]{devellisScaleDevelopmentTheory2022} provides a four-step framework: (1) \textit{item creation}, (2) \textit{item evaluation and weight choice}, (3) \textit{validity examination}, and (4) \textit{reliability examination}.
We adapt Liu et al.'s approach~\cite{liuITrustEvalFrameworkSoftware2022}, using \ac{GQM} to derive measurements from trust factors and \ac{AHP} to determine empirical weights.


\subsection{Item Creation via \ac{GQM}}
\label{sec:item-creation}

While DeVellis and Thorpe~\cite{devellisScaleDevelopmentTheory2022} emphasize social science surveys, we require automated metrics for \ac{OSS} analysis.
Thus we adopt Basili and Weiss's \ac{GQM} method from software engineering~\cite{basiliMethodologyCollectingValid1984}.

\ac{GQM} proceeds in three steps:

\begin{enumerate}
\item \textbf{Goals}: Define measurement objectives across five dimensions: object, purpose, quality focus, viewpoint, context~\cite[p.\ 18]{anacletoGQMhandbookOverviewGQMplans}. ``Without goals, one is likely to obtain data in which either incomplete patterns or no patterns are discernible''~\cite[p.\ 3]{basiliMethodologyCollectingValid1984}.

\item \textbf{Questions}: Derive 1+ questions per goal to enable quantitative analysis, forming an intermediate abstraction layer between subjective goals and objective metrics~\cite[pp.\ 23--24]{anacletoGQMhandbookOverviewGQMplans,basiliMethodologyCollectingValid1984}.

\item \textbf{Metrics}: Select metrics with defined unit, scale, range, and collection protocol~\cite[p.\ 27]{anacletoGQMhandbookOverviewGQMplans}. Follow van Solingen et al.'s guidelines: prefer existing data sources, objective over subjective measures, and reliability-supporting metrics~\cite[p.\ 7]{vansolingenGoalQuestionMetric2002}.
\end{enumerate}

Applying \ac{GQM} to our trust factors yields 12 concrete, automatable measurements (Table~\ref{tab:questions}), each becoming a \ac{DCI} index item.

\subsection{Item Evaluation and Weight Choice}\label{sec:ahp}

After \ac{GQM} item creation, developers must select final index items and assign weights~\cite[pp.\ 204--212]{devellisScaleDevelopmentTheory2022}.
Regression-based weighting requires robust ground truth data, which trustworthiness lacks due to its subjective, multi-faceted nature.
Instead, since \ac{DCI} addresses the \textit{dependency selection decision problem}, we adopt Chen et al.~\cite{chenModelMethodTrustworthiness2010} and Liu et al.'s~\cite{liuITrustEvalFrameworkSoftware2022} \ac{AHP} approach.

\ac{AHP} structures decisions as a three-tier hierarchy (goal $\rightarrow$ criteria $\rightarrow$ alternatives)~\cite[p.\ 2]{saatyModelsMethodsConcepts2012}.
It computes empirical relative importance weights for each criterion with respect to the goal, enabling informed selection among dependency alternatives.



\subsection{Validity Examination}\label{sec:validity}

Index validity requires accurately representing the target composite variable~\cite[p.\ 71]{devellisScaleDevelopmentTheory2022}.
DeVellis~\cite[p.\ 216]{devellisScaleDevelopmentTheory2022} recommends criterion validation for formative indices, demonstrating relatedness to a relevant outcome (criterion) via correlation or mean differences~\cite[p.\ 933]{horstmannCriterionValidity2020}.

We use \textit{concurrent validity}, comparing \ac{DCI} scores to simultaneously collected ground truth~\cite[p.\ 933]{horstmannCriterionValidity2020,piedmontCriterionValidity2023}.
Since direct trustworthiness measurement is infeasible, we adopt OpenSSF Scorecard scores as proxy ground truth-capable of identifying vulnerable projects with 78\% accuracy~\cite{magillReportFindsOpenSSF2022}.
Pearson correlation and linear regression assess \ac{DCI}-Scorecard alignment (Sec.~\ref{sec:eval}).

\subsection{Reliability Examination}\label{sec:reliability}

Reliability requires an index to stably reflect its target variable under unchanged conditions~\cite[p.\ 33]{devellisScaleDevelopmentTheory2022}.
Internal consistency (suitable for reflective scales) does not apply to formative indices~\cite[p.\ 36,218]{devellisScaleDevelopmentTheory2022}.

Instead, DeVellis~\cite[p.\ 218]{devellisScaleDevelopmentTheory2022} recommends test-retest stability.
We compute \ac{DCI} scores twice on the same packages and verify score invariance (Sec.~\ref{sec:eval}).


\section{Methodology}
\label{sec:methodology}

\ac{DCI} is a composite formative index combining indicators that collectively represent \ac{OSS} library trustworthiness~\cite[p.\ 186]{devellisScaleDevelopmentTheory2022}.
While subjective trust factors lack established causal theory - warranting skepticism~\cite[p.\ 188]{devellisScaleDevelopmentTheory2022} - a simple formative index suits developers' dependency selection needs.

\ac{DCI} items derive from Hou and Jansen's \ac{SLR}~\cite{houSystematicLiteratureReview2023}, with measurements selected from prior research and \ac{AHP}-derived empirical weights (Sec.~\ref{sec:ahp}).

Trust factor selection follows \ac{GQM} item creation (Sec.~\ref{sec:item-creation}).
Using Anacleto et al.'s five dimensions~\cite[p.\ 18]{anacletoGQMhandbookOverviewGQMplans}, we define the measurement goal:

\begin{quote}
\textit{The goal of \ac{DCI} is to evaluate the trustworthiness of a third-party \ac{OSS} library whenever a software developer selects dependencies for their current project.}
\end{quote}

\ac{GQM} requires deriving questions from the goal using literature-derived trust factors. We adapt nine factors from Hou and Jansen~\cite[pp.\ 20--27]{houSystematicLiteratureReview2023}, categorized per Hongwei et al.~\cite{hongweitaoAttributesOrientedSoftware2022} as attribute-based (software artifacts) or process-based (development practices).

\begin{table}[ht!]
    \begin{center}
    \caption{Selected trust factors adapted from~\cite[pp.\ 20--27]{houSystematicLiteratureReview2023}.}
    \label{tab:selected-factors}
    \begin{tabular}{ll}
        \toprule
        \multirow{4}{*}{\textbf{Attribute-based}} & Security \textbf{(SV)} \\
        & Source Code Quality \textbf{(CQ)} \\
        & Documentation Completeness \textbf{(DO)} \\
        & License Declaration \textbf{(SA)} \\
        \midrule
        \multirow{5}{*}{\textbf{Process-based}} & Development Process Quality \textbf{(QP)} \\
        & Project Health \textbf{(PH)} \\
        & Release Cadence \textbf{(VD1)} \\
        & Dependency Management \textbf{(VD2)} \\
        & Reputation \textbf{(RE)} \\
        \bottomrule
    \end{tabular}
    \end{center}
\end{table}

We exclude behavior-based approaches, as executing untrusted dependencies risks malicious code exposure.
Attribute-based factors (e.g., cyclomatic complexity, code smells) analyze artifacts directly; process-based factors (e.g., release frequency, issue closure) examine GitHub repository practices.

\textbf{SV} consolidates vulnerabilities and security antipatterns per~\cite{houSystematicLiteratureReview2023}, excluding vulnerability frequency (reserved for validation).
This narrow definition ensures specificity, avoiding overlap with general quality metrics.
SV directly addresses trustor expectations of intended behavior, where documented vulnerabilities and insecure coding practices (SonarQube) violate core trustworthiness.

\textbf{CQ} is the second factor. While source code quality and documentation quality are conventionally assessed collectively as
components of overall software quality~\cite[p. 2-2]{mccallFactorsSoftwareQuality1977}, this study delineates them as distinct trust factors.
This separation is implemented to establish more granular and specific trust factors, thereby avoiding overly broad categorizations that lack intuitive meaning for developers utilizing the index.
A classic model of software quality is provided by McCall~\cite[p. 3-5]{mccallFactorsSoftwareQuality1977}. The quality factors as defined by~\cite{mccallFactorsSoftwareQuality1977}
are shown in Table~\ref{tab:mccall-quality}. This study employs one factor from each category that is applicable within the context of \ac{OSS} libraries,
specifically: reliability, maintainability, and reusability. Reusability represents a special case, as it is inherently satisfied in \ac{OSS} contexts and therefore does not require explicit evaluation.

\begin{table}[ht!]
    \centering
    \caption{Software quality factors adapted from \cite[Fig. 3.1-1]{mccallFactorsSoftwareQuality1977}}
    \label{tab:mccall-quality}
    \begin{tabular}{lll}
        \toprule
        Product Operation & Product Revision & Product Transition \\
        \midrule
        Correctness & Maintainability & Portability \\
        Reliability & Flexibility & Reusability \\
        Efficiency & Testability & Interoperability \\
        Integrity &  &  \\
        Usability &  & \\
        \bottomrule
    \end{tabular}
\end{table}

An additional attribute-based trust factor is \textbf{DO}. While McCall et al.~\cite{mccallFactorsSoftwareQuality1977} define
documentation as a component of software, they do not classify documentation as a software quality factor per se,
but rather as an artifact of the development process. This study adopts a different approach.
Given that the trust factors employed in this research must be amenable to automated evaluation,
and considering the absence of standardized methodologies for identifying documentation based on a given package
and the inherent complexity of correlating specific documentation sections with corresponding source code,
this study evaluates the completeness of code comments. Consequently, this trust factor is categorized as attribute-based rather than process-based.
This approach matches the trust factor definition by Hou and Jansen~\cite{houSystematicLiteratureReview2023}.

The last attribute-based trust factor is \textit{License Declaration}, which is derived from \textbf{SA}.
The existence of a license is important because it allows the developer to use the library in their application.
Since this is the only aspect of \textbf{SA} that can be automatically measured, this study redefines \textbf{SA} to mean \textit{License Declaration}.
Furthermore, Hou and Jansen~\cite[pp. 26-27]{houSystematicLiteratureReview2023} define cost as a trust factor.
However, cost considerations are not applicable in the context of \ac{OSS}, as open source libraries are provided without charge to users.
Therefore, \textbf{CO} is not included in \ac{DCI}.
The exemption of payment is contingent upon the presence of a declared license,
as libraries without explicit licensing cannot be legally utilized
without written permission from the copyright holder.
Unauthorized usage of such libraries would expose users to potential legal liability and litigation risk.
Therefore, for such a software to be trustworthy, a license must be declared.

\begin{table*}[ht!]
    \centering
    \caption{List of questions derived from trust factors.}
    \label{tab:questions}
    \begin{tabular}{cp{7.5cm}}
    \toprule
        Trust Factor & Question \\
        \midrule
        \multirow{2}{*}{Security \textbf{(SV)} } 
        & \textbf{S1:} How many vulnerabilities are present in the code and how severe are they?\\
        & \textbf{S2:} How many security antipatterns are contained in the code? \\
        \hdashline
        \multirow{2}{*}{Source Code Quality \textbf{(CQ)} } 
        & \textbf{C1:} How many reliability issues are contained in the code? \\
        & \textbf{C2:} How much technical debt exists in the codebase? \\
        \hdashline
         Documentation \textbf{(DO)} & \textbf{D1:} How well is the code commented? \\
         \hdashline
         Structural Assurance \textbf{(SA)} & \textbf{L1:} Is a license declared in the project? \\
         \hdashline
        \multirow{2}{*}{Development Process Quality \textbf{(QP)} } 
        & \textbf{P1} Does the project use continuous integration? \\
        & \textbf{P2} What is the test coverage? \\
        \hdashline
         Project Health \textbf{(PH)}  & \textbf{H1:} How diverse are the maintainers of the project? \\
         \hdashline
         Release Cadence \textbf{(VD1)}  & \textbf{R1:} How frequently does the project release new versions? \\
         Dependency Management \textbf{(VD2)}  & \textbf{M1:} Does the project use dependency management tools? \\
         \hdashline
         Reputation \textbf{(RE)}  & \textbf{T1:} How popular is the project? \\
    \bottomrule
    \end{tabular}
\end{table*}

The first process-based trust factor is \textbf{QP}.
Following Hou and Jansen~\cite{houSystematicLiteratureReview2023}, software quality as a trust factor is split into two separate factors that together influence quality, yet are assessed using distinct metrics. This study defines development process quality as a \say{set of framework and umbrella activities, actions, and work tasks}~\cite[p. 2]{singhImpactSoftwareDevelopment2016} that are used to produce a software product. These activities directly impact the quality of the resulting software~\cite[p. 3]{singhImpactSoftwareDevelopment2016}.
Therefore, it is essential to assess whether a software library is developed utilizing appropriate supporting activities and tooling, including the implementation of \ac{CI}, fuzzing tools, and \acp{ASAT}.
The adoption of automated testing frameworks enhances software product quality by increasing fault detection rates and
delivering more reliable testing outcomes, thereby contributing to overall software dependability~\cite[p. 37]{dudekulamohammadrafiBenefitsLimitationsAutomated2012}.

A further process-based trust factor is \textbf{PH}. Project health is defined as the ability of an \ac{OSS} project to
\say{stay viable and maintained over time without interruption or weakening}~\cite[p. 1]{linakerHowCharacterizeHealth2022}.
This metric is of critical importance given that \ac{OSS} software has become an integral component of contemporary computing infrastructure~\cite[p. 1]{gogginsOpenSourceCommunity2021},
and unmaintained projects are unable to address security vulnerabilities~\cite[p. 1]{linakerHowCharacterizeHealth2022},
thereby presenting risks of widespread systemic disruptions.

A factor that is closely related to \textbf{PH} is \textit{Release Cadence} \textbf{(VD1)}.
This work splits the trust factor \textbf{VD} into two separate factors.
\cite[p. 25]{houSystematicLiteratureReview2023} states that \say{[t]he distribution, installation, and updating of software packages
or components also impact software trustworthiness, especially the integrity and reliability of the software}.
This definition does not differentiate between the distribution of the software that is being evaluated and its transitive dependencies.
Therefore, this work splits \textbf{VD} into the trust factors \textbf{VD1} and \textbf{VD2}.
\textbf{VD1} is defined as \say{the measure of time between software releases, both internal and external}~\cite[p. 1]{kilicKeepBallRolling2023}.
While project health attempts to estimate the long-term survivability of an \ac{OSS} project,
release cadence evaluates whether the project has already become unmaintained. To minimize vulnerability risks within a project,
dependencies must be current~\cite[p. 25]{houSystematicLiteratureReview2023}, as unmaintained projects fail to provide essential security updates.

The second trust factor derived from \textbf{VD} is \textit{Dependency Management} \textbf{(VD2)}.
Most software libraries utilize dependencies of their own~\cite{houSystematicLiteratureReview2023}.
Consequently, all trust factors applicable to the primary software recursively extend to every dependency within the dependency tree.
Given the computational expense of evaluating trust metrics for each transitive dependency, it is necessary to verify whether the library employs dependency management software.
This verification ensures that the library receives critical security updates and bug fixes~\cite{kulaTrustingLibraryStudy2015}.

The last process-based trust factor is \textbf{RE}. There are two types of reputation: software reputation and software producer reputation~\cite[p. 25]{houSystematicLiteratureReview2023}.
This work focuses on software reputation, because this is the most common type that developers use when evaluating a library~\cite[p. 4]{duanMeasuringSupplyChain2020}.
Software reputation is defined as \say{information cues about the source of the code such as the number of reviews, the origin (e.g., website, colleagues), or the number of
users}~\cite[p. 2]{alarconTrustPerceptionsMetadata2020}. This study defines popularity as a sub-factor of reputation in contrast to Hou and Jansen~\cite{houSystematicLiteratureReview2023}.
Alarcon et al.~\cite[p. 11]{alarconTrustPerceptionsMetadata2020} show that reputation is a crucial attribute that developers rely on heavily when choosing a trustworthy software.

Based on the above selection of trust factors, questions can be derived using the  \ac{GQM} methodology.
Table~\ref{tab:questions} shows those questions. 13 questions are derived from the nine trust factors.
The questions reflect the factor definitions as presented above.

\subsection{Weight Estimation}\label{sec:weights}

As detailed in Section~\ref{sec:ahp}, this study employs the \ac{AHP} method to determine the weights for the indices
A group of 10 developers with different backgrounds and ages was asked to complete the \ac{AHP} survey.
The age range was 25 to 40, and the programming experience ranged from Junior to Senior Software Engineers.
The programming languages in which the participants are proficient are Python, Java, and C++.
All the respondents were men.
They were instructed to rate the relative importance of each of the 72 pairs. With nine trust factors, the pairwise comparison matrix contains $9\times9 = 81$ entries. Excluding the nine diagonal entries (self-comparisons fixed to 1.0), participants effectively provided 72 comparison judgements per questionnaire (cf. Table~\ref{tab:ahp_matrix}).
The study utilizes a tool provided by Goepel~\cite{goepelImplementationOnlineSoftware2018} for creating the questionnaire and calculating the results.
The results are presented in Table~\ref{tab:weights} and Appendix~\ref{a:ahpsurvey}.

\begin{table}[ht!]
    \centering
    \caption{\ac{AHP}-derived weights for \ac{DCI} trust factors.}
    \label{tab:weights}
    \begin{tabular}{lr}
        \toprule
        \textbf{Trust Factor} & \textbf{Weight} \\
        \midrule
        SV (Security) & 0.277 \\
        CQ (Source Code Quality) & 0.164 \\
        PH (Project Health) & 0.140 \\
        VD2 (Dependency Management) & 0.096 \\
        DO (Documentation) & 0.092 \\
        QP (Development Process) & 0.073 \\
        SA (License Declaration) & 0.072 \\
        VD1 (Release Cadence) & 0.049 \\
        RE (Reputation) & 0.037 \\
        \midrule
        \multicolumn{1}{r}{\textit{Total}} & \textit{1.000} \\
        \bottomrule
    \end{tabular}
\end{table}

The results show that most respondents rate security as the most important factor when considering third-party software, followed by source code quality and project health.
These three factors account for approximately 60\% of the total weight.
The two factors with the lowest importance are release cadence and reputation.

The \ac{AHP} sample should be interpreted as an exploratory convenience sample rather than as a representative sample of \ac{OSS} developers. Accordingly, the weights reported in Table~\ref{tab:weights} are treated as provisional empirical weights. Their purpose in this study is to provide a transparent and reproducible initial weighting scheme for the pilot implementation, not to establish a definitive hierarchy of trust factors. A larger follow-up study should include participants with different professional roles, levels of experience, industry backgrounds, geographic regions, and demographic characteristics.

Since the \ac{AHP} survey was exploratory and involved only 10 participants, the weights should be regarded as an initial configuration of the index. The present study does not include a weight-sensitivity analysis. In particular, it does not establish whether package rankings, \ac{DCI} distributions, or the correlation with OpenSSF Scorecard remain stable under alternative weighting schemes. Such an analysis requires access to the package-level measurement data and will be included in future validation work (cf. Sec. \ref{sec:future-work}).

\subsection{Measurement Selection}

\ac{DCI} combines technical and non-technical attribute measurements as proposed by Immonen and Palviainen~\cite{immonenTrustworthinessEvaluationTesting2007}.
However, in contrast to their research, \ac{DCI} not only infers non-technical measurements from the software architecture
but also directly measures them using publicly available data on GitHub.
This score should be easily interpretable, providing the software developer with clear guidance on whether to use or avoid a particular library.
According to the \ac{GQM} methodology, the measurements are derived based on the questions defined in Table~\ref{tab:questions}.

\begin{table}[ht!]
    \centering
    \caption{\ac{DCI} measurements – \ac{GQM} questions (Q), units, types, calculations, and normalization.}
    \label{tab:measurements}
    \begin{tabular}{c p{2.3cm} p{2.4cm} c p{3.1cm} c}
        \toprule
        Q & Measurement & Unit & Type & Calculation & Normalization \\
        \midrule
        S1 & Vulnerability Density & vulnerabilities/\newline\acs{KLoC} & cont. & vulnerabilities/\acs{KLoC} & $1 - x/27$ \\
        S2 & Security Issue Density & security issues/\acs{KLoC} & cont. & security issues/\acs{KLoC} & $1 - x/27$ \\
        \hdashline
        C1 & Bug Density & bugs/\acs{KLoC} & cont. & bugs/\acs{KLoC} & $1 - x/27$ \\
        C2 & Code Smell Density & code smells/\newline\acs{KLoC} & cont. & code smells/\acs{KLoC} & $1 - x/27$ \\
        \hdashline
        D1 & Comment Density & ratio & cont. & comment lines/\acs{SLoC} & $2x$ \\
        \hdashline
        L1 & License Declaration & Boolean & Binary & approved license present & $x$ \\
        \hdashline
        P1 & \ac{CI} & Boolean & Binary & known \ac{CI} tool present & $x$ \\
        P2 & Code Coverage & ratio & cont. & executed \acs{SLoC}/total \acs{SLoC} & $x$ \\
        \hdashline
        H1 & Bus Factor & maintainers & cont. & \cite[Alg.~1]{avelinoNovelApproachEstimating2016} & $x/10$ \\
        \hdashline
        R1 & Release Frequency & releases/year & cont. & releases / age (in days) $\cdot 365$ & $x/36$ \\
        \hdashline
        M1 & Dependency Management & Boolean & Binary & known dependency management tool present & $x$ \\
        \hdashline
        T1 & Popularity & GitHub stars & cont. & GitHub stars & $x/3800$ \\
        \bottomrule
    \end{tabular}
\end{table}


Table~\ref{tab:measurements} shows those measurements, presenting the measurement chosen for each question.
It also provides a definition of the calculation for each measurement as well as the scoring methodology.

Boolean measurements are represented as 0 or 1. Continuous measurements are normalized to the interval ($[0,1]$) using the transformations shown in the final column. The labels “cont.” and “binary” describe the measurement representation; they do not imply that the resulting normalization bounds are empirically calibrated for all ecosystems.

The normalization constants in Table~\ref{tab:measurements} are implementation-specific reference values rather than universal quality thresholds. Several values originate from studies of other programming languages or project populations and may not transfer directly to Python packages. They were used in the pilot implementation to map heterogeneous measurements to a common normalized interval.

This choice has two consequences. First, a score of 1 does not necessarily indicate an absolute optimum, and a score of 0 does not necessarily indicate complete untrustworthiness. Second, the resulting \ac{DCI} values depend partly on the selected reference bounds. The current study therefore treats the numerical scale as provisional and does not claim that the normalization is calibrated for all \ac{OSS} ecosystems.

\ac{DCI} uses four types of defect densities to answer questions \textbf{S1}, \textbf{S2}, \textbf{C1}, and \textbf{C2}.
It measures the number of defects that occur per 1,000 lines of code. This study distinguishes between vulnerabilities, security issues,
bugs, and code smells to answer the questions specifically. Shah et al.~\cite[p. 414]{shahOverviewSoftwareDefect2012} provide a mean estimate of
7.47 defects per 1000 lines of code with an upper bound of 27~\cite[Fig. 7]{shahOverviewSoftwareDefect2012}. The upper bound is used as a divisor for calculating the score based on a defect density measurement.
Since this is only an approximation, this study will evaluate this methodology in Sec.~\ref{sec:eval}.
The measurement for \textbf{D1}, comment density, is already a number bounded by 0 and 1. However, Arafat and Riehle~\cite[Fig. 3]{arafatCommentDensityOpen2009}
show that no project exceeds a comment density of more than 50\%. Therefore, the density value is doubled to map it to the required bounds.
\textbf{L1}, \textbf{P1}, \textbf{P2}, and \textbf{M1} directly utilize the measurement result since they are already within the correct range.
The value of measurement \textbf{H1} is divided by 10. The value of the bus factor algorithm used in this work rarely exceeds 10~\cite[Fig. 5]{avelinoNovelApproachEstimating2016}.
Therefore, 10 is used as an upper bound.
Joshi and Chimalakonda~\cite[Tab. 1]{joshiRapidReleaseDatasetProjects2019} calculate the mean for distinct releases of GitHub projects to be $42.38$, and the mean of the total
age to be $884.16$ days. Based on the numbers, the average number of releases per days is $0.049$. Since the standard deviation is not available for this
data, an upper bound of $2 * 0.049 * 365 = 35.81$ releases per year is estimated.
Lastly, the upper bound of the popularity of Java applications on GitHub uses the third quartile as calculated by~\cite[Fig. 4]{borgesUnderstandingFactorsThat2016}.

\subsection{System Architecture \& Implementation}
\label{ss:implementation}

\begin{figure}[ht!]
    \centering
    \includegraphics[width=0.97\linewidth]{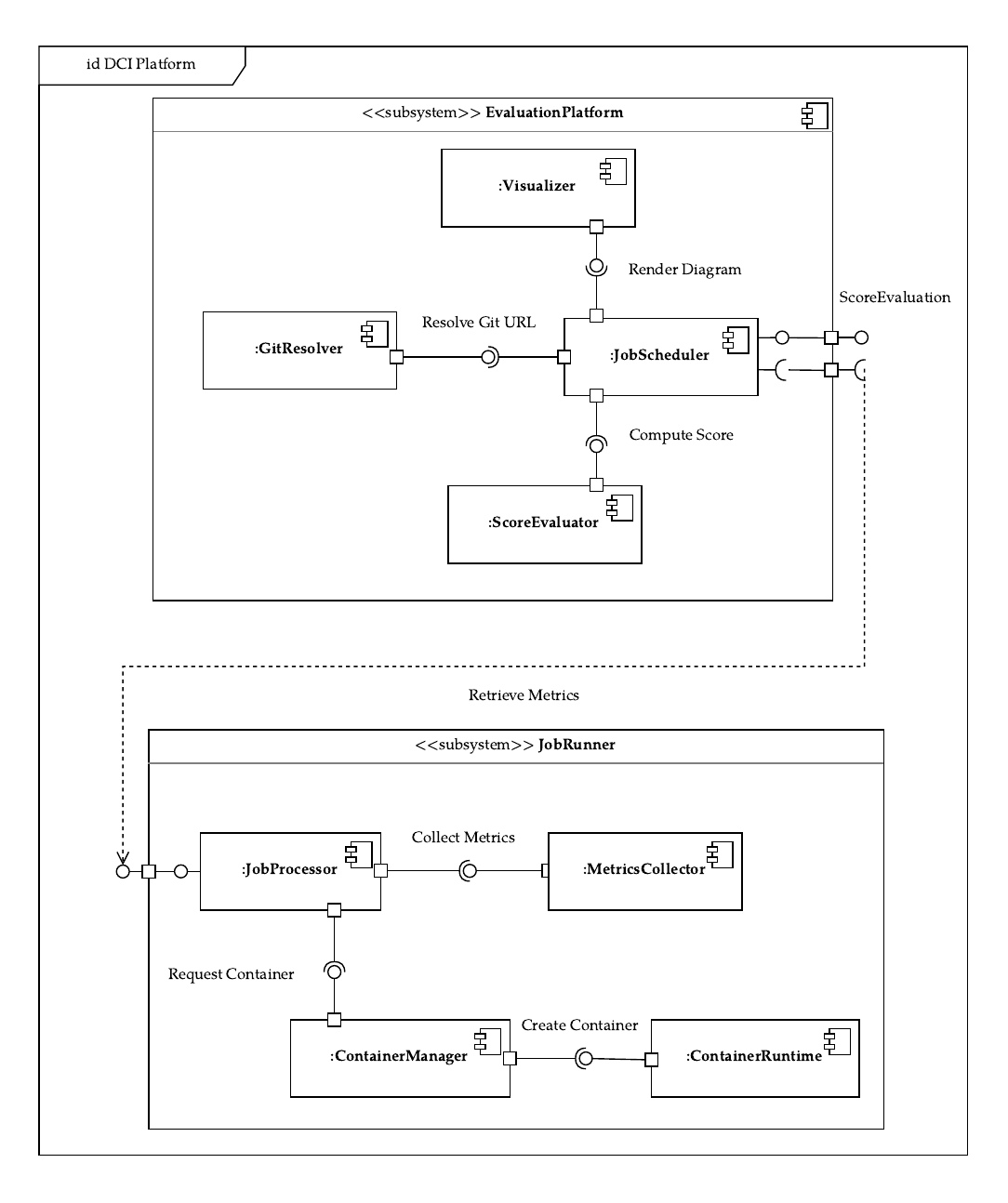}
    \caption{Component diagram of system architecture showing the subsystems Evaluation Platform and the JobRunner.}
    \label{fig:platform}
\end{figure}

At its core, \ac{DCI} is a metrics collection platform. The two main tasks are collecting the metrics, and calculating the index. In terms of implementation, \ac{DCI} comprises two subsystems for metric collection and index computation (Fig.~\ref{fig:platform}) - both are publicly available at \url{https://doi.org/10.5281/zenodo.19349535} with added readme for configuration and running:

\begin{itemize}
\item \textbf{Evaluation Platform}: 
  \begin{itemize}
  \item Web interface submits package details
  \item Handover of package to GitHub repository
  \item Schedules analysis jobs asynchronously
  \end{itemize}

\item \textbf{JobRunner} (containerized):
  \begin{itemize}
  \item Clones repository in isolated environment
  \item Runs SonarQube static analysis + GitHub analysis
  \item Computes raw measurements and returns results
  \end{itemize}
\end{itemize}

The implementation artifact contains the source code and scripts required to operate the platform and reproduce the intended analysis workflow. Full reproduction of the historical numerical results additionally depends on the exact package versions, database exports, external \ac{API} responses, tool versions, configuration files, and service availability used during the original experiment. These dependencies should be considered when interpreting the reproducibility claim.

Users submit package specifications (PURL/name) via Angular to a JakartaEE backend, where the \textit{JobScheduler} resolves them to GitHub URLs using the \textit{GitResolver} component. The \textit{JobScheduler} then dispatches jobs to available JobRunners. Each JobRunner operates in Docker isolation: the \textit{JobProcessor} requests containers from the \textit{ContainerManager}, which instructs the \textit{ContainerRuntime} to spin up secure environments.
Inside these containers, the MetricsCollector builds the untrusted package and executes SonarQube static analysis alongside GitHub \ac{API} scrapers to gather raw metrics.

Completed metrics return to the Evaluation Platform, where the \textit{ScoreEvaluator} computes the weighted \ac{DCI} and the \textit{Visualizer} generates petal diagrams.
This asynchronous containerized design handles SonarQube timeouts ($\leq$30min) while ensuring security through isolation.

\subsubsection{Evaluation Platform Design}

\begin{figure}[ht!]
    \centering
    \includegraphics[width=0.97\linewidth]{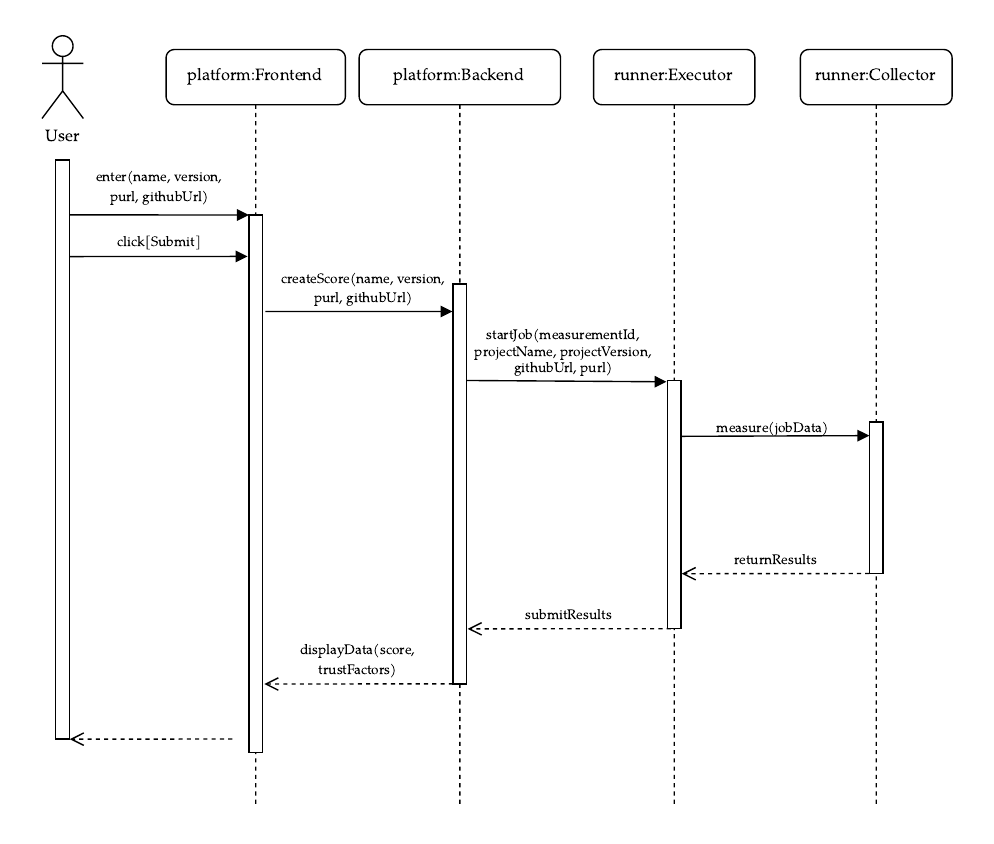}
    \caption{Sequence diagram showing the information flow through the different components.}
    \label{fig:dcisequence}
\end{figure}

The evaluation platform supports both standalone usage and \ac{CI}/\ac{CD} integration through a modular REST API backend (JakartaEE) and optional Angular frontend.
This separation enables backend-only deployments when the UI is unnecessary, establishing clear testable boundaries. Figure \ref{fig:dcisequence} presents the interaction between the different components. 

\begin{figure}[ht!]
    \centering
    \includegraphics[width=0.7\linewidth]{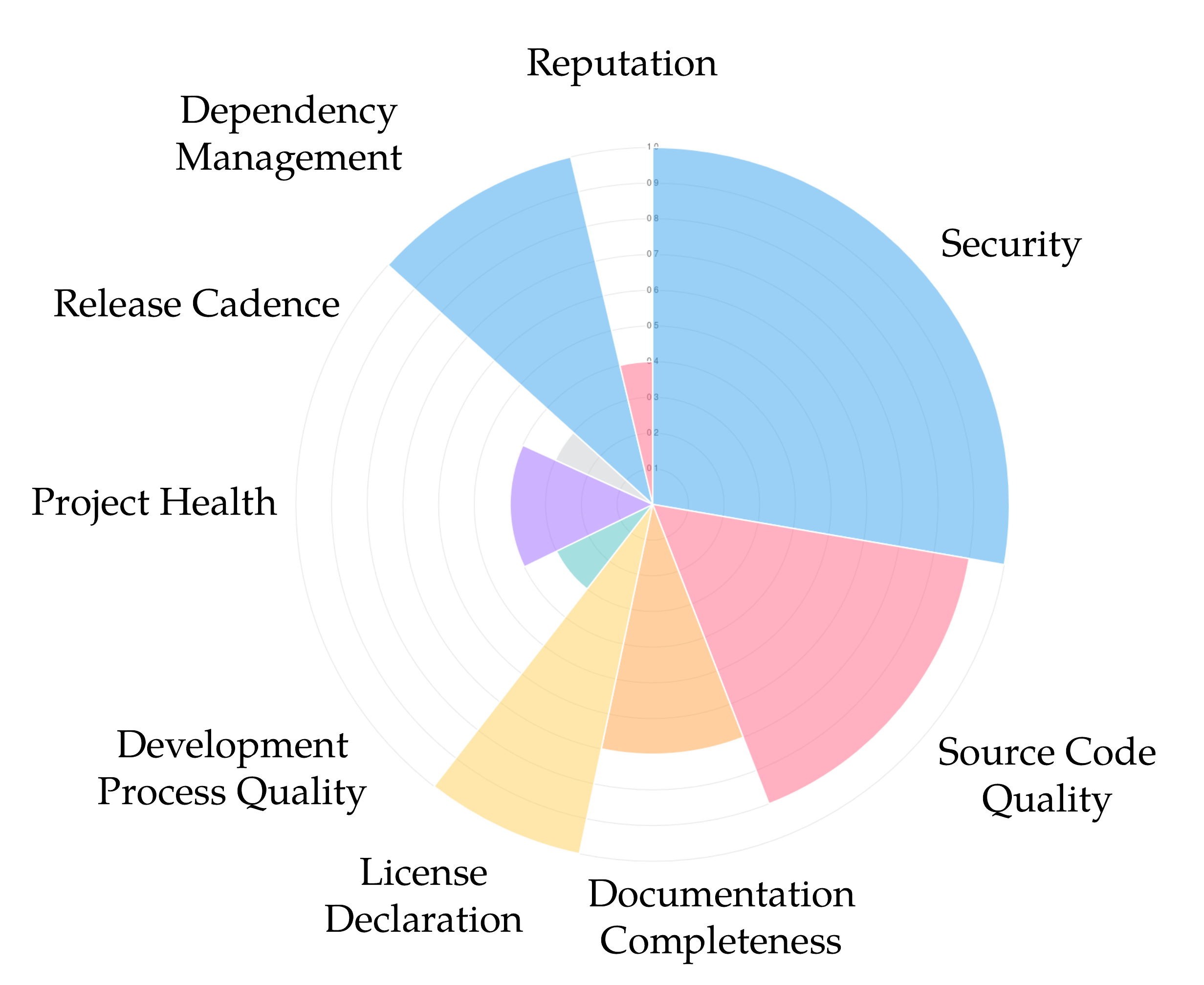}
    \caption{Petal chart of the evaluation platform showing the value of each trust factor and its weight.}
    \label{fig:petal}
\end{figure}

\textbf{Frontend} (Angular) provides five pages: greeter, login (JWT-based, stored in localStorage to avoid CSRF), home (score listing), createScore (package input: name/version/PURL/ GitHub URL), and scoreDetail (metrics + petal visualization).
The petal diagram (Fig. \ref{fig:petal}) proportionally represents \ac{AHP} weights (Sec.~\ref{sec:ahp}) - each trust factor's segment angle is calculated via $360^\circ \times \text{weight}$.

\textbf{Backend} provides \ac{JWT} authentication, WebSocket job scheduling to distributed runners, and REST API with database persistence (PostgreSQL). The key workflow is as follows:
\begin{itemize}
\item User submits package $\rightarrow$ creates score entity, selects available runner, schedules analysis
\item Runner completes $\rightarrow$ metrics stored, \ac{DCI} score computed automatically
\end{itemize}

The asynchronous WebSocket design and PostgreSQL persistence enable long-running SonarQube scans while maintaining responsive user experience.

\subsubsection{JobRunner}

The JobRunner receives measurement jobs via WebSocket from the Evaluation Platform and executes them in Docker-isolated containers. It comprises an \textit{Executor} (WebSocket orchestration + container management) and \textit{Collector} (actual metric gathering).

A JobRunner executor authenticates via \ac{JWT}, establishes a persistent WebSocket connection, and upon job receipt schedules isolated container execution. The \textit{ContainerManager} ensures current collector images (built from Dockerfile if needed) with proper environment configuration. The \textit{MetricsCollector} performs five core tasks within the container:
\begin{itemize}
\item \textbf{Source preparation}: \texttt{git clone} + tag matching + \texttt{git checkout} for exact version
\item \textbf{SonarQube analysis} (S2,C1,C2,D1,P2): Auto-detects Java build tool (Gradle/Maven), configures SonarScanner, polls measures \ac{API}
\item \textbf{Vulnerabilities} (S1): Queries \ac{OSV} via PURL+version
\item \textbf{Bus factor} (H1): Executes Avelino's algorithm~\cite{avelinoNovelApproachEstimating2016}
\item \textbf{Process/project} (L1,P1,M1,R1,T1): OpenSSF Scorecard + GitHub \ac{API} + Libraries.io metadata
\end{itemize}

Raw metrics return via \acs{HTTP} PUT to trigger \ac{DCI} computation. Container isolation protects against malicious packages.

Three major challenges emerged during \ac{DCI} development: 

\textit{Build automation:} Automating Java builds proved difficult - approximately 50\% of projects fail to compile automatically due to \ac{JDK} incompatibilities, non-default build parameters, and configuration errors~\cite{hassanAutomaticBuildingJava2017}. This explains why most static analysis research targets interpreted languages like Python.

\textit{Package metadata quality:} Automatic source URL resolution failed due to incomplete/incorrect metadata.
Only 58\% of PyPI and 46\% of npm packages have valid source URLs; common issues include missing URLs and inaccessible GitHub links~\cite{tsakpinisAnalyzingAccessibilityGitHub2024}.
This undermines open source assumptions and hinders codebase research.

\textit{Lack of standardization:} Jakarta EE \ac{JWT} configuration (OpenLiberty) required trial-and-error across overlapping \acp{API} (JSON Web Token, mpJWT).
Standards lag creates proprietary extensions, reducing portability and documentation.

The Java implementation experience should be interpreted as an engineering feasibility observation rather than as an empirical comparison between Java and Python. The observed build failures were caused by a combination of \ac{JDK} incompatibilities, non-default build parameters, and project-specific configuration requirements. Because the failed and successful projects may differ systematically, the successful subset cannot be assumed to be representative of Java projects generally. These practical hurdles motivated shifting evaluation to Python packages, where builds and metadata proved more reliable.




        

\section{Evaluation}
\label{sec:eval}

The empirical evaluation reported in this paper is limited to Python packages, as Java compilation and project configuration created a high failure rate during data collection. Python was therefore selected for the pilot evaluation because the analyzed packages could be processed without a compilation step. This decision improved execution feasibility but limits the generalizability of the findings across programming languages and package ecosystems.

Therefore, to validate \ac{DCI}'s criterion validity and test-retest reliability, we collected scores for popular Python packages using the evaluation platform. Although initially designed for Java (Gradle/Maven support), pervasive build failures led to evaluating Python packages, where the same platform succeeded. The implementation of \ac{DCI}, as detailed in Sec.~\ref{ss:implementation}, focuses on the analysis of Java applications. As described, during experimentation with the software, it was discovered that this leads to a high likelihood of a failed analysis, because SonarQube requires Java projects to be compiled. Compiling Java projects is very difficult to automate uniformly, and the compilation process can take up to an hour, depending on the hardware. This issue made it very difficult to collect the data necessary for the evaluation. Therefore, the focus of this project was shifted to Python libraries, as they are interpreted and don't require compilation.

We retrieved a list of 100 packages from \textit{Libraries.io\footnote{\url{https://libraries.io}}}. This package list is used to collect the metadata that is necessary to compute the \ac{DCI} scores. The Curl package was used to query the \textit{Libraries.io} \ac{API} with the command as shown here:

\begin{lstlisting}[language=bash]
curl 'https://libraries.io/api/search' \
  --get \
  --data-urlencode 'api_key=<API_KEY>' \
  --data-urlencode 'per_page=100' \
  --data-urlencode 'languages=Python' \
  --data-urlencode 'platforms=Pypi' \
  --data-urlencode 'sort=dependents_count'
\end{lstlisting}


This \acs{HTTP} request queries the \ac{API} to return 100 packages written in Python and distributed through the Python Packaging Index.
To ensure that the packages are actual libraries and not complete applications, the list is by the most depended-upon packages.
This package selection has the downside that the packages will likely be of high quality resulting in a narrow distribution of scores.
The complete list of analyzed packages (name, GitHub URL, version) is provided as supplementary material.

For Python, a small program performs the collection of \ac{DCI} scores. This program logs into the \ac{DCI} platform of Sec.~\ref{ss:implementation} and requests the analysis of a package from the list of test packages every 30 minutes to avoid reaching the \ac{API} request limits of GitHub. After finishing the analysis of all packages, the results were dumped from the database in SQL format. To process the results easily, the SQL data was converted into a CSV file.

To collect the Scorecard scores, the website \textit{open/source/insights}\footnote{\url{https://deps.dev/}} provides pre-computed Scorecard values for most popular open-source libraries. These values were retrieved in a loop from the \ac{API} and written to a CSV file. The Scorecard values are divided by 10 to match the value range of \ac{DCI}.

Out of the 100 analyzed packages (including \textit{numpy, requests, pandas, pytest, Django, Flask, scikit-learn}, and \textit{transformers}), eight could not be evaluated. Three of the eight failed packages were missing a GitHub URL. SonarQube could not analyze the other five. Out of the 100 Scorecard scores requested, 90 scores could be retrieved, as the used service does not provide scores for all possible packages. Because there is only a small overlap between packages missing from \ac{DCI} and Scorecard, the number of packages available for evaluation is 85.

\subsection{Scores}
\label{ss:scores}

The \ac{DCI} scores have a mean value of \textbf{0.665} with a standard deviation of \textbf{0.087}.
This elevated mean aligns with the expectation that, since these are popular packages, the scores will likely be high.
The Scorecard scores have a mean of \textbf{0.622} with a standard deviation of \textbf{0.130}.

\begin{figure}[ht!]
    \centering
    \includegraphics[width=.8\linewidth]{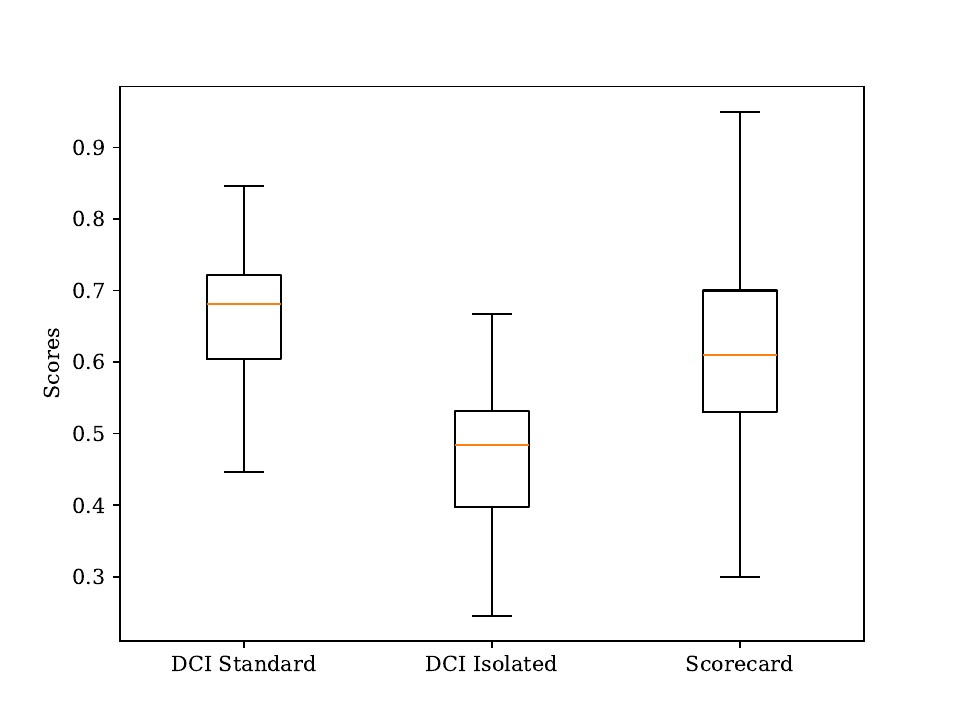}
    \caption{Comparison of the value distributions of different score types.}
    \label{fig:boxplot}
\end{figure}

Figure~\ref{fig:boxplot} compares the means of these two score types. It also presents a third type: \textit{\ac{DCI} Isolated}. This type is an equal-weighted version of \ac{DCI}, which only contains measurements not included in the Scorecard values.
The excluded measurements are dependency management, license declaration, and the use of \ac{CI} tools.
The value range of Scorecard scores contains both the ranges of standard and isolated \ac{DCI} scores.
Standard \ac{DCI} has the highest mean, followed by Scorecard and isolated \ac{DCI}.

\begin{figure}[ht!]
    \centering
    \includegraphics[width=.8\linewidth]{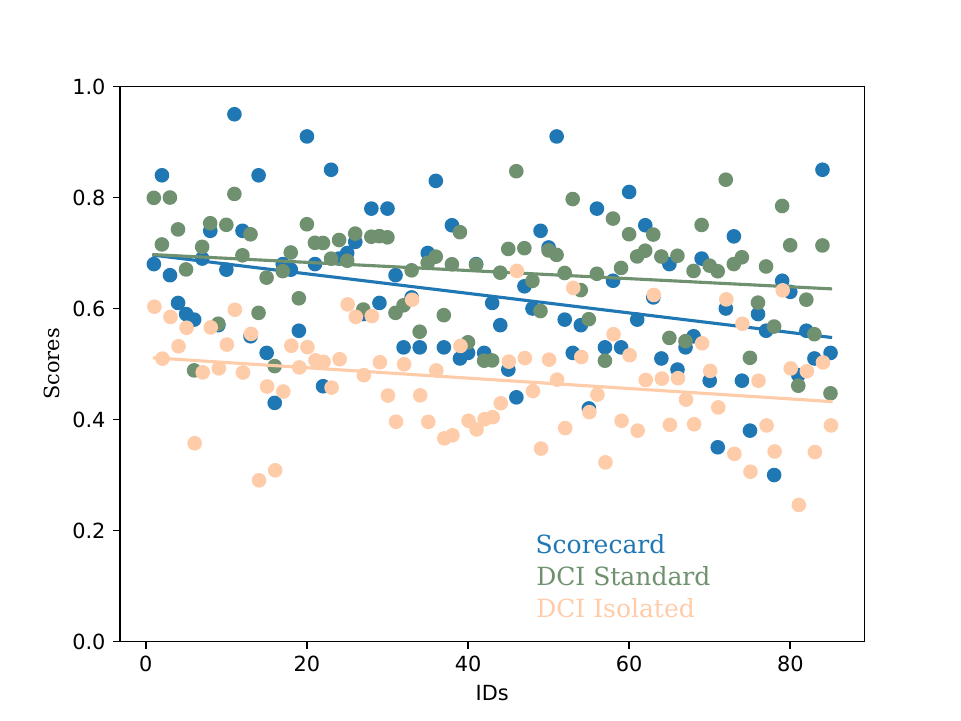}
    \caption{Comparison of the value distributions of different score types with trend lines.}
    \label{fig:dci-distribution}
\end{figure}

Figure~\ref{fig:dci-distribution} shows the distribution and trend line of the three
score types. All three trend lines are linear and reflect the approximately linear distribution of all score values.
The linearity allows the computation of Pearson coefficients to measure the correlation between these scores.
Note the similarity of the slope of the trend lines. The values in these diagrams are ordered by
ID on the $x$-axis. Since the IDs are themselves ordered by decreasing number of dependent packages,
the trend lines indicated a positive correlation between package popularity and package quality.
Furthermore, these plots indicate that all three types of plots should be correlated,
which a calculation of the correlations confirms. There is a moderate, but statistically significant
correlation between standard \ac{DCI} values and Scorecard values, with a correlation coefficient of \textbf{0.396},
and a $p$-value of \textbf{0.0002}.
The correlation between isolated \ac{DCI} scores and Scorecard is less statistically significant and weaker,
with a correlation of \textbf{0.228} and a p-value of \textbf{0.036}.

\begin{table}[ht!]
    \centering
    \caption{Pearson coefficients between each trust factor and the \ac{DCI} score ($n=85$).}
    \label{tab:correlations}
    \begin{tabular}{lrr}
        \toprule
        \textbf{Trust Factor} & \textbf{$r$} & \textbf{$p$} \\
        \midrule
        VD2 (Dependency Management) & 0.760 & $<.001$ \\
        QP (Development Process) & 0.602 & $<.001$ \\
        PH (Project Health) & 0.556 & $<.001$ \\
        RE (Reputation) & 0.518 & $<.001$ \\
        CQ (Source Code) & 0.392 & $0.0002$ \\
        SA (License Declaration) & 0.345 & $0.001$ \\
        VD1 (Release Cadence) & 0.324 & $0.003$ \\
        DO (Documentation) & 0.307 & $0.004$ \\
        SV (Security) & 0.033 & $0.767$ \\
        \bottomrule
    \end{tabular}
\end{table}

Table~\ref{tab:correlations} presents the Pearson coefficients between each trust factor and the \ac{DCI} score.
The factor \textbf{SV} is particularly noticeable. This factor is not significantly correlated to the final score. Therefore, security does not explain the observed variation in the \ac{DCI} scores in this dataset. The non-significant association for \textbf{SV} must be interpreted in light of the measurement distribution. Both security measurements were constant at $1.0$ for the evaluated package versions. A constant factor cannot explain variation in the composite score, regardless of its assigned \ac{AHP} weight. Thus, the observed ($r=0.033$) and ($p=0.767$) indicate that security did not contribute observable score variation in this dataset, but they do not indicate that security is unimportant.

This distinction is particularly important because the \ac{AHP} survey assigned Security the largest factor weight. The discrepancy between conceptual importance and empirical influence illustrates why factor weights alone do not guarantee discriminative power: the factor must also be measured with a valid and sufficiently variable indicator.

All other trust factors are correlated with the final score, with a $p$-value indicating statistical significance.
The trust factors with the strongest correlation are: \textbf{QP}, \textbf{PH}, \textbf{VD2},
and \textbf{RE}. The causes of these correlations are investigated in the next section,
as the detailed measurement data requires analysis.


\subsection{Measurements}\label{subsec:analysis-measurements}

\begin{figure}[ht!]
    \centering
    \includegraphics[width=.8\linewidth]{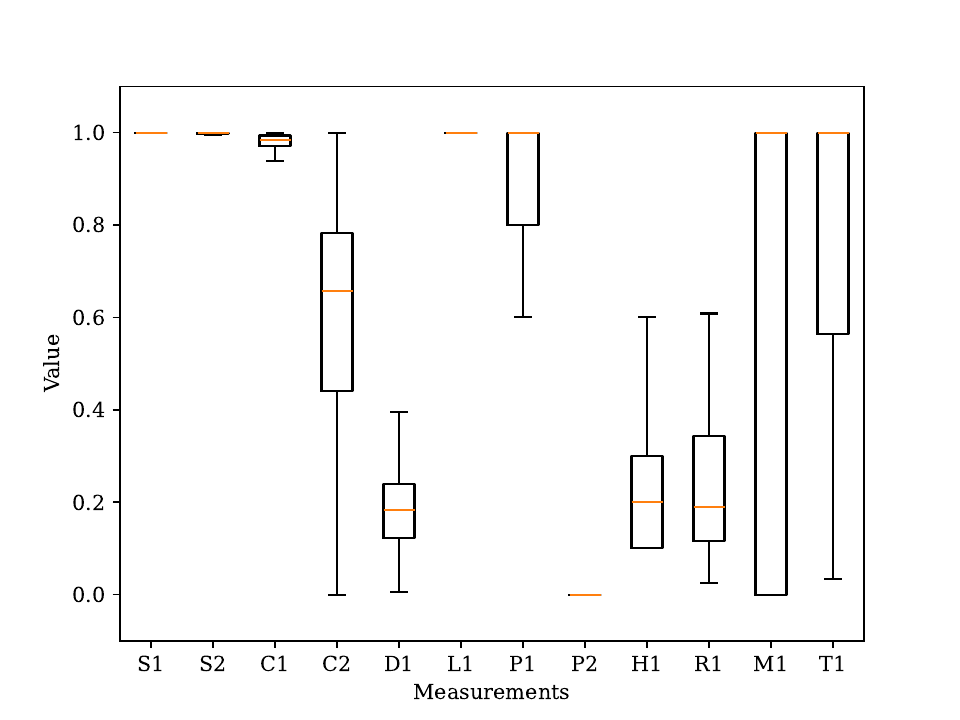}
    \caption{Value distribution of \ac{DCI} measurements. The measurements form three empirical classes in the evaluated dataset. \textbf{S1, S2, L1}, and \textbf{P2} are constant and therefore non-discriminative in this sample. M1 is binary and occupies both extreme values.\textbf{ C1, C2, D1, P1, H1, R1}, and \textbf{T1} show non-constant distributions. These classes explain why some factors contribute little or no variance to the reported \ac{DCI} scores, whereas binary or highly variable measurements may have a stronger influence.}\label{fig:measurement-dist}
\end{figure}

To explain the correlation between each trust factor and the \ac{DCI} score,
we analyzed the value distribution of each measurement, as these represent the raw data.
Figure~\ref{fig:measurement-dist} shows the minimum, maximum, median, first, and third quartiles of the measurements.

Using Figure~\ref{fig:measurement-dist}, the measurements can be grouped into three classes.
Measurements \textbf{S1}, \textbf{S2}, \textbf{L1}, and \textbf{P2} have a constant value of
either $1.0$ or $0.0$, forming the first class. The second class contains one measurement, \textbf{M1},
which has a box that fills the value range from $0.0$ to $1.0$. This shape is caused by the measurement flipping between
these two values. Lastly, \textbf{C1}, \textbf{C2}, \textbf{D1}, \textbf{P1}, \textbf{H1}, \textbf{R1}, and \textbf{T1} represent
the class of regular measurements as these measurements have a regular value distribution.

The first class of constant values explains the lack of influence that the trust factor \textbf{SV} has on the score.
For this trust factor, both measurements are constant across all measured packages.
Only the most recent version of 100 popular Python packages were measured.
Therefore, none of the measured packages contained known vulnerabilities. Furthermore,
it demonstrates that these projects prioritize secure coding practices,
as no security issues were found in the code during static analysis.
These facts explain the lack of influence that this trust factor has.
\textbf{L1} has a constant value of $1.0$. This value is expected because all the measured packages are open-source.
For a project to be open-source, it has to declare a license. This factor will only have a value of $0.0$
for malicious packages. It is therefore still important, but irrelevant to the analysis.
In the present implementation, \textbf{P2} did not provide a valid measurement of code coverage for the evaluated Python packages. It therefore has to be treated as unavailable rather than as evidence of zero coverage. The constant value of $0.0$ resulted from SonarQube’s inability to obtain valid coverage data for these projects. Treating an unavailable measurement as a numerical zero can reduce the interpretability of the composite score and may bias the \textbf{QP} factor.

The second class contains only measurement \textbf{M1}. This measurement, which, like \textbf{L1} and \textbf{P2},
is a binary measurement. Since there is a large number of packages for both extreme values, this measurement, and as
a result, the trust factor \textbf{VD2} has the most significant influence on the score.
This influence affects the expressiveness of the \ac{DCI} score and could be mitigated by reducing its weight.

The last class is the measurements, which have regular values. This class can be further subdivided into three groups.
\textbf{C1} and \textbf{P1} have consistently large values across the dataset. \textbf{C1} shows that
these popular projects don't release new versions with a large number of bugs. This lack of bugs is as expected.
\textbf{P1} shows the prevalence of \ac{CI} practices in these popular projects which helps explain
the lack of bugs, as \ac{CI} is typically used to run regression tests.
Next are \textbf{D1}, \textbf{H1}, and \textbf{R1}. These measurements have mostly low values.
These consistently low values might be the result of insufficiently high upper bounds. Therefore, these measurements should be tested across
a larger data set, and the upper bound should be reevaluated. Lastly, \textbf{C2} and \textbf{T1} have a large
variance in the data. The effect of \textbf{C2} on the \ac{DCI} score is lowered due to the averaging
with \textbf{C1} in the trust factor. The larger variance of \textbf{T1} explains the high correlation of \textbf{RE}.

Upon examining the individual measurements, five stand out with an average score of approximately \textbf{1.0}.
These are bug density, license availability, security issue density, code smell density, and vulnerability density.
Another anomaly is the constant zero value of the code coverage measurement.
This constant zero value is due to SonarQube's inability to accurately estimate this value.
This issue can only be remedied by manually configuring the test code discovery, which is beyond the scope of this experiment.

The measurement distributions show that the current dataset does not provide an equally informative test of all \ac{DCI} factors. \textbf{S1}, \textbf{S2}, and \textbf{L1} are constant at 1.0, while \textbf{P2} is constant at 0.0 because code coverage was not successfully extracted. These measurements therefore do not discriminate between packages in this evaluation.

\textbf{M1} is binary and changes between 0 and 1 for a substantial part of the dataset. Consequently, it contributes disproportionately to score variation relative to measurements with continuous distributions. This observation does not imply that dependency management is intrinsically more important than the other factors; it reflects the interaction between the assigned weight, the binary measurement design, and the composition of the evaluated package sample.

The resulting \ac{DCI} values should therefore be interpreted as scores from the current implementation and dataset, not as fully calibrated estimates of dependency trustworthiness. In particular, the absence of variation in \textbf{S1} and \textbf{S2} means that the present experiment cannot evaluate the discriminative contribution of the security factor.

\subsection{Validity Examination}\label{sec:analysis-validity-examination}

The goal of the validity examination is to show that \ac{DCI} and Scorecard have a statistically significant correlation, as outlined in Section~\ref{sec:validity}.
The results of computing the Pearson correlation are an $r$ of \textbf{0.38}, with a p-value of \textbf{0.004}.
Therefore, it can be concluded that \ac{DCI} performs similarly to Scorecard.
A slightly weaker form of this correlation can also be shown with the isolated version of \ac{DCI}
described in Section~\ref{ss:scores}.
As a result, \ac{DCI} is a valid index for measuring the trustworthiness of open-source libraries.
However, given that the actual value remains fundamentally unknowable due to its inherently subjective nature,
it is important to acknowledge that the values generated by \ac{DCI} constitute estimates rather than definitive measurements.

\subsection{Reliability Examination}\label{sec:analysis-reliability--examination}

According to Section~\ref{sec:reliability}, the only method applicable to the reliability examination of \ac{DCI}
is the test-retest methodology. Therefore, half of the packages were tested twice. The research halved the number of packages for the reliability analysis to shorten the processing time. 

The test-retest experiment demonstrates computational repeatability for the subset of packages evaluated twice under unchanged conditions and exact package versions. It does not establish temporal stability under changing repository metadata, newly disclosed vulnerabilities, updated Scorecard data, or changed normalization and weighting schemes. We therefore interpret the result as evidence of repeatability of the implementation under controlled inputs, not as definitive evidence that \ac{DCI} reliably measures dependency trustworthiness in all contexts.


\section{Conclusion}\label{sec:conclusion}

This work aimed to develop a methodology for constructing an open-source library trust index.
To achieve this goal, the research involved, first, investigating the relevant trust factors through a literature review.
Second, it entailed determining the relative weights of these trust factors using a survey.
Third, it included identifying the corresponding software metrics for each trust factor.
Based on these metrics, the research proceeded by developing a software architecture.
Finally, it encompassed investigating the validity and reliability of the developed index.

\subsection{Key Findings}\label{subsec:key-findings}

The research started with the establishment of a definition of trust and trustworthiness.
It established that trustworthiness is a property of an object or subject, and
that trust is an action between a subject and a different subject or object~\cite{becerraTrustworthinessRiskTransfer2008}.
Trust was defined as a subjective action~\cite{houSystematicLiteratureReview2023}, and the trustworthiness of software
as the degree to which it conforms to both functional and non-functional requirements~\cite{amorosoProcessorientedMethodologyAssessing1994}.
Furthermore, the research established that trust is transitive within a specific domain~\cite{liuTrustTransitivityComplex2011}.
Therefore, the trustworthiness of a software producer influences the trustworthiness of the software itself.

Based on these definitions, this work established nine trust factors using a comparative literature review.
These factors are: \textit{Security}, \textit{Source Code Quality}, \textit{Documentation Completeness},
\textit{License Declaration}, \textit{Development Process Quality}, \textit{Project Health}, \textit{Release Cadence},
\textit{Dependency Management}, and \textit{Reputation}.
To capture the subjectivity of trustworthiness, a survey using the \ac{AHP} method was conducted.
The participants of the survey were a group of software developers, as they are the target demographic for \ac{DCI}.
After performing \ac{AHP} the research found, that the most important trust factor for this work is security,
closely followed by source code quality and project health. Based on these trust factors,
a set of twelve quantitative measures was derived. These mesures were designed to be automatically
collectable.

To evaluate the validity of \ac{DCI}, the calculation of \ac{DCI} for 100 Python packages was performed.
These packages are the 100 most depended upon packages in the Python Packaging Index.
Of these 100 packages, 92 could be successfully measured.
It was found that both \ac{DCI} and Scorecard scores are linearly distributed.
The means of both \ac{DCI} and Scorecard scores are similar, with a difference of $0.043$.
After calculating the Pearson correlation coefficient for each trust factor with the resulting \ac{DCI} scores,
it was found that Dependency Management (\textbf{VD2}) has the most significant influence on the score,
followed closely by Development Process Quality (\textbf{QP}), Project Health (\textbf{PH}), and Reputation (\textbf{RE}).

The pilot evaluation provides preliminary evidence that the proposed framework can be computed repeatedly under unchanged inputs and that its scores show moderate association with OpenSSF Scorecard in the evaluated sample. These findings do not establish \ac{DCI} as a generally valid security measure or as a complete representation of supply-chain risk.

The results showed a moderate correlation between the standard \ac{DCI} and Scorecard, which is statistically significant.
Because alternative weighting schemes were not evaluated, the current results do not demonstrate that the observed package rankings or the correlation with OpenSSF Scorecard are invariant to the \ac{DCI} outcome.
It also demonstrated that even after removing measurements, which are present in both \ac{DCI} and Scorecard, there is still a correlation.
After applying the test-retest methodology, the research demonstrated that \ac{DCI} scores can be reliably repeated.

\subsection{Limitations}\label{sec:limitations}

The limitations of this work can be grouped into three categories.
These limitations include the validity of the index,
its applicability to different contexts, and the limitations of the measurements.
The most important limitation is the limited validity of \ac{DCI}.
As trustworthiness is fundamentally subjective in nature, its actual value cannot be objectively determined.
Therefore, a proxy indicator has to be used to establish validity.
However, the validity of the proxy indicator itself can be questioned.
The best estimation of the true validity of \ac{DCI} could be established
through its successful utilization in detecting supply chain attacks.
This should be addressed in a larger-scale future study.

The \ac{AHP}-based weighting scheme is another important limitation. The sample consisted of only ten homogeneous respondents. Although individual pairwise comparisons and consistency ratios are provided in Appendix A, the sample size and composition limit the generalizability of the aggregated weights. Future work should replicate the survey with a larger and more diverse participant population and should investigate whether the factor hierarchy remains stable across developer and security-related roles.

The limited applicability of \ac{DCI} to all programming languages is a result
of the use of a literature review to establish upper bounds for the measurements.
The upper bounds of \textbf{S1}, \textbf{S2}, \textbf{C1}, \textbf{C2}
are based on a study of the C, C++, and Java programming languages~\cite{shahOverviewSoftwareDefect2012}.
Furthermore, this study utilized GitHub stars as a measure of a library's popularity.
Therefore, this measurement does not apply to projects hosted on other source code management systems. The current dataset does also not permit conclusions about the behavior of \ac{DCI} for unpopular, abandoned, newly created, malicious, or historically vulnerable packages. In particular, the absence of variation in the security measurements is partly a consequence of the package-selection strategy. A more demanding evaluation should use stratified sampling and include lower-quality and less-maintained projects.

The normalization bounds are not ecosystem-neutral. The defect-density reference value is based on literature involving C, C++, and Java projects, while the popularity and release-frequency bounds were derived from other project populations. Their use for Python packages is therefore an engineering approximation rather than an empirically validated calibration. Future work should derive robust reference distributions from larger, language-specific datasets and should evaluate alternative transformations such as percentile-based, robust, or rank-based scaling.

The current evaluation cannot establish the security sensitivity of \ac{DCI}. The analyzed packages were popular and generally mature, and none of the evaluated package versions produced variation in the vulnerability-density or security-issue-density measurements. Consequently, the high \ac{AHP} weight assigned to security did not translate into a measurable influence on the \ac{DCI} distribution. This is a limitation of the dataset and measurement execution, not evidence that security is unimportant.

Similarly, the constant license value and defective code-coverage measurement reduce the effective dimensionality of the index. A future implementation should validate the variance and availability of each measurement before score aggregation. Measurements that are constant, unavailable, or technically invalid should either be excluded with corresponding weight redistribution or reported separately rather than being silently included in the composite score.

Lastly, three measurements did not produce satisfactory results.
The measurements \textbf{S1}, and \textbf{S2} had a constant value of $1.0$.
The choice of packages in the dataset caused this.
More data on older and less popular packages is necessary to study the effects of these measurements properly.
Measurement \textbf{P2} had a constant value of $0.0$ because the program was unable to measure
the code coverage of the analyzed projects. This issue is a limitation of the implementation and should be
addressed to enable further analysis of the index.

\subsection{Outlook}\label{sec:outlook}

This work envisions two usage scenarios for \ac{DCI}.
The first scenario involves using it as an editor plugin. In this scenario,
the user installs a plugin into their preferred editor or \ac{IDE}.
This plugin scans the currently open project to determine its language and build tool.
If the user opens the configuration file of the build tool, the plugin will show a \ac{DCI} score next
to every dependency entry as an inline hint. Furthermore, if the score is below a certain threshold,
the dependency should be marked with a warning.

The second scenario involves its use as part of a \ac{CI} pipeline. In this scenario,
\ac{DCI} is used to evaluate the trustworthiness of dependencies not on the user's client device
but as part of a git repository. If a user pushes code to a feature branch and creates a merge request,
a \ac{CI} runner will look up dependencies in the code and calculate the \ac{DCI} score for each dependency it finds.
If a dependency receives a score under a certain configurable threshold, the runner will create an issue
in the repository and block merging until the issue is resolved.

\subsection{Future Work}\label{sec:future-work}

This work provided a first set of quantitative measurements of the trustworthiness of software libraries.
However, there are still interesting questions left that can be studied.
First, the change in \ac{DCI} scores for a specific
package can be studied over time. This data could be used to
capture the effect of specific development changes on a software's trustworthiness.

Future work should also perform a systematic weight-sensitivity analysis. This analysis should vary the factor weights while preserving their sum, recompute package-level \ac{DCI} scores, compare rank-order stability, and assess the effect on correlations with external reference measures such as OpenSSF Scorecard. Suitable analyses could include equal weighting, leave-one-factor-out variants, perturbation of individual weights, and alternative weights obtained from larger \ac{AHP} samples.

Issue-tracking systems provide additional information that is not currently represented in \ac{DCI}. Potential indicators include issue closure time, unresolved-issue age, issue reopening rate, the proportion of issues linked to releases, security-issue response time, and the evolution of issue backlogs. These indicators could refine \textbf{PH} and \textbf{QP}.

Reliability-growth models could provide a complementary longitudinal view by relating defect discovery and correction processes to project releases. However, such models require consistent historical observations, well-defined failure or defect categories, and sufficient time-series data. Their integration should therefore be evaluated separately rather than treated as a direct replacement for the current static measurements.

Furthermore, different programming languages could be analyzed to study the trustworthiness differences between
them.
There are also improvements to the software implementation that could be performed.
The collection time of the measurements that use static analysis
could be improved if these measurements could be collected without the need
to compile the library first. This enhancement would also enable the evaluation of more programming languages using this index,
as many packages can't be measured due to compilation issues.
Additionally, future research should focus on investigating a correlation between the number of
bugs and code smells and the occurrence of detected security vulnerabilities in software. This improvement would enable
the definition of better reference numbers for the trust factor source code quality, which
in turn would improve the validity of that factor.

\newpage

\section*{Abbreviations}
The following acronyms were used in this work:
\begin{acronym}
\itemsep=-7pt
\acro{AHP}{Analytic Hierarchy Process}
\acro{API}{Application Programming Interface}
\acro{ASAT}{Automatic Static Analysis Tool}
\acro{CD}{Continuous Development}
\acro{CI}{Continuous Integration}
\acro{CRUD}{Create, Read, Update, Delete}
\acro{CSRF}{Cross-site Request Forgery}
\acro{CSV}{Comma-seperated Values}
\acro{DAO}{Data Access Object}
\acro{DCI}{Dependency Confidence Index}
\acro{DOM}{Document Object Model}
\acro{GQM}{Goal/Question/Metric}
\acro{HTML}{Hypertext Markup Language}
\acro{HTTP}{Hypertext Tranfer Protocol}
\acro{IDE}{Integrated Development Environment}
\acro{JDK}{JAVA Development Kit}
\acro{JSON}{JavaScript Object Notation}
\acro{JWT}{JSON Web Token}
\acro{KLoC}{Thousands of Lines of Code}
\acro{LLM}{Large Language Model}
\acro{MCDM}{Multicriteria Decision Making}
\acro{OpenSSF}{Open Source Security Foundation}
\acro{OSI}{Open Source Initiative}
\acro{OSS}{Open Source Software}
\acro{OSV}{Google Open Source Vulnerability database}
\acro{PURL}{Package URL}
\acro{REST}{Representational State Transfer}
\acro{SBOM}{Software Bill of Materials}
\acro{SLoC}{Source Lines of Code}
\acro{SLR}{Systematic Literature Review}
\acro{SPA}{Single Page Application}
\acro{UML2}{Unified Modeling Language Version 2}
\acro{URL}{Unified Resource Locator}
\acro{VND}{Version Number Delta}
\end{acronym}

\section*{Appendix}
\label{a:ahpsurvey}

\begin{table}[H]
\centering
\small
\renewcommand{\arraystretch}{1.15}
\caption{Weights for Security and Source Code Quality.}
\label{tab:weights-1}
\begin{tabular}{|c|c|c|}
\hline
\textbf{Participant} & \textbf{Security} & \textbf{Source Code Quality} \\
\hline
p1  & 0.231586 & 0.034320 \\
p2  & 0.039091 & 0.235128 \\
p3  & 0.431908 & 0.170008 \\
p4  & 0.428215 & 0.165068 \\
p5  & 0.280231 & 0.167666 \\
p6  & 0.217185 & 0.356564 \\
p7  & 0.242980 & 0.093728 \\
p8  & 0.162033 & 0.106137 \\
p9  & 0.324231 & 0.147638 \\
p10 & 0.418406 & 0.092713 \\
\hline
\end{tabular}
\end{table}

\begin{table}[H]
\centering
\small
\renewcommand{\arraystretch}{1.15}
\caption{Weights for Documentation Completeness and License Declaration.}
\label{tab:weights-2}
\begin{tabular}{|c|c|c|}
\hline
\textbf{Participant} & \textbf{Documentation Completeness} & \textbf{License Declaration} \\
\hline
p1  & 0.023237 & 0.310371 \\
p2  & 0.383565 & 0.017438 \\
p3  & 0.022441 & 0.081686 \\
p4  & 0.086000 & 0.088638 \\
p5  & 0.043840 & 0.042398 \\
p6  & 0.113446 & 0.018032 \\
p7  & 0.098596 & 0.051700 \\
p8  & 0.137881 & 0.072823 \\
p9  & 0.039790 & 0.212601 \\
p10 & 0.169195 & 0.020627 \\
\hline
\end{tabular}
\end{table}

\begin{table}[H]
\centering
\small
\renewcommand{\arraystretch}{1.15}
\caption{Weights for Development Process Quality and Project Health.}
\label{tab:weights-3}
\begin{tabular}{|c|c|c|}
\hline
\textbf{Participant} & \textbf{Development Process Quality} & \textbf{Project Health} \\
\hline
p1  & 0.033996 & 0.160636 \\
p2  & 0.071594 & 0.070026 \\
p3  & 0.035184 & 0.133019 \\
p4  & 0.061706 & 0.074049 \\
p5  & 0.103599 & 0.192161 \\
p6  & 0.125022 & 0.059076 \\
p7  & 0.047430 & 0.217991 \\
p8  & 0.040896 & 0.217011 \\
p9  & 0.051381 & 0.117627 \\
p10 & 0.065670 & 0.070220 \\
\hline
\end{tabular}

\end{table}

\begin{table}[H]
\centering
\small
\renewcommand{\arraystretch}{1.15}
\caption{Weights for Release Cadence and Dependency Management.}
\label{tab:weights-4}
\begin{tabular}{|c|c|c|}
\hline
\textbf{Participant} & \textbf{Release Cadence} & \textbf{Dependency Management} \\
\hline
p1  & 0.044706 & 0.085502 \\
p2  & 0.055964 & 0.085700 \\
p3  & 0.047054 & 0.067988 \\
p4  & 0.044641 & 0.030440 \\
p5  & 0.028367 & 0.099924 \\
p6  & 0.021088 & 0.073255 \\
p7  & 0.046630 & 0.164838 \\
p8  & 0.031640 & 0.137063 \\
p9  & 0.025484 & 0.066435 \\
p10 & 0.061463 & 0.063347 \\
\hline
\end{tabular}

\end{table}

\begin{table}[H]
\centering
\small
\renewcommand{\arraystretch}{1.15}
\caption{Weights for Reputation and Consistency Ratio.}
\label{tab:weights-5}
\begin{tabular}{|c|c|c|}
\hline
\textbf{Participant} & \textbf{Reputation} & \textbf{CR} \\
\hline
p1  & 0.075648 & 0.074241 \\
p2  & 0.041494 & 0.157657 \\
p3  & 0.010712 & 0.195414 \\
p4  & 0.021243 & 0.144621 \\
p5  & 0.041813 & 0.226036 \\
p6  & 0.016333 & 0.339019 \\
p7  & 0.036107 & 0.116396 \\
p8  & 0.094517 & 0.050998 \\
p9  & 0.014813 & 0.138138 \\
p10 & 0.038359 & 0.068102 \\
\hline
\end{tabular}
\end{table}

\begin{table}[H]
\scriptsize
    \centering
    \caption{\ac{AHP} pairwise comparison matrix.}
    \label{tab:ahp_matrix}
    \renewcommand{\arraystretch}{1.2}
    \begin{tabular}{ccccccccc}
        \midrule
        1.000000 & 1.857888 & 3.485451 & 3.801230 & 4.057553 & 2.630717 & 4.427319 & 2.501653 & 5.829850 \\
        0.538245 & 1.000000 & 2.159439 & 2.452591 & 2.784299 & 0.967683 & 3.579789 & 1.558193 & 3.717347 \\
        0.286907 & 0.463083 & 1.000000 & 1.705556 & 1.374109 & 0.722741 & 1.621948 & 0.889798 & 2.666081 \\
        0.263073 & 0.407732 & 0.586319 & 1.000000 & 1.075678 & 0.472577 & 1.390389 & 0.848945 & 2.452591 \\
        0.246454 & 0.359157 & 0.727744 & 0.929646 & 1.000000 & 0.565312 & 1.834630 & 0.698827 & 2.497717 \\
        0.380125 & 1.033396 & 1.383622 & 2.116056 & 1.768936 & 1.000000 & 3.415430 & 1.450205 & 3.801230 \\
        0.225870 & 0.279346 & 0.616542 & 0.719223 & 0.545069 & 0.292789 & 1.000000 & 0.512497 & 1.517601 \\
        0.399736 & 0.641769 & 1.123850 & 1.177933 & 1.430969 & 0.689558 & 1.951232 & 1.000000 & 2.010677 \\
        0.171531 & 0.269009 & 0.375082 & 0.407732 & 0.400366 & 0.263073 & 0.658935 & 0.497345 & 1.000000 \\
        \bottomrule
    \end{tabular}
\end{table}

\newpage

\bibliographystyle{ieeetr}
\bibliography{bibliography}
\end{document}